\documentclass[prd,nofootinbib,preprint,12pt]{revtex4}

\usepackage[english]{babel}
\usepackage{amsmath,amssymb}
\usepackage[dvips]{graphicx}

\allowdisplaybreaks

\newcommand{\obs}{\text{obs}}

\newcommand{\ie}{\textit{i.e.}}
\newcommand{\eg}{\textit{e.g.}}

\newcommand{\pdf}{\textrm{p.d.f.}}
\newcommand{\dof}{\textrm{d.o.f.}}
\newcommand{\sgof}{\textrm{SG}}
\newcommand{\pgof}{\textrm{PG}}

\newcommand{\by}{ {\rm\bf y} }
\newcommand{\bx}{ {\rm\bf x} }
\newcommand{\bv}{ {\rm\bf v} }
\newcommand{\bmu} {\mbox{\boldmath$\mu$}}
\newcommand{\bnu} {\mbox{\boldmath$\nu$}}
\newcommand{\btheta}{\mbox{\boldmath$\theta$}}

\newcommand{\LT}{\left}
\newcommand{\RT}{\right}

\newcommand{\CL}{\mathop{\rm CL}\nolimits}

\newcommand{\dms}{\Delta m^2_\mathrm{sol}}
\newcommand{\dma}{\Delta m^2_\mathrm{atm}}

\newtheorem{propos}{Proposition}

\begin{document}

\preprint{IFIC/03-13}
\preprint{TUM-HEP-506/03}

\vspace*{2cm}
\title{Testing the statistical compatibility of independent data sets}

\author{M.~Maltoni} \email{maltoni@ific.uv.es}
\affiliation{Instituto de F\'{\i}sica Corpuscular --
  C.S.I.C./Universitat de Val{\`e}ncia \\
  Edificio Institutos de Paterna, Apt 22085,
  E--46071 Valencia, Spain}

\author{T.~Schwetz} \email{schwetz@ph.tum.de}
\affiliation{Institut f{\"u}r Theoretische Physik, Physik Department\\
  Technische Universit{\"a}t M{\"u}nchen,
  James--Franck--Str., D--85748 Garching, Germany
  \vspace*{2 cm}
  }

\begin{abstract}
    We discuss a goodness-of-fit method which tests the compatibility
    between statistically independent data sets. The method gives
    sensible results even in cases where the $\chi^2$-minima of the
    individual data sets are very low or when several parameters are
    fitted to a large number of data points. In particular, it avoids
    the problem that a possible disagreement between data sets becomes
    diluted by data points which are insensitive to the crucial
    parameters. A formal derivation of the probability distribution
    function for the proposed test statistic is given, based on
    standard theorems of statistics. The application of the method is
    illustrated on data from neutrino oscillation experiments, and its
    complementarity to the standard goodness-of-fit is discussed.
\end{abstract}

\maketitle

\section{Introduction}

The essence of any scientific progress is the comparison of theoretical
predictions to experimental data. Statistics provides the scientist with
so-called {\it goodness-of-fit} tests, which allow to obtain well defined
probability statements about the agreement of a theory with data. The by far
most popular goodness-of-fit test dates back to 1900, when K.~Pearson
identified the minimum of a $\chi^2$-function as a powerful tool to evaluate
the quality of the fit~\cite{pearson}.
However, it is known that the Pearson $\chi^2_\mathrm{min}$ test is
not very restrictive in global analyses, where data from different
experiments with a large number of data points are compared to a
theory depending on many parameters.  The reason for this is that in
such a case a given parameter is often constrained only by a small
subset of the data. If the rest of the data (which can contain many
data points) are reasonably fitted, a possible problem in the fit of
the given parameter is completely washed out by the large amount of
data points. A discussion of this problem in various contexts can be
found \eg\ in Refs.~\cite{collins,Maltoni:2001bc,solGof}.

To evade this problem a modification of the original
$\chi^2_\mathrm{min}$ test was proposed in Ref.~\cite{Maltoni:2002xd}
to evaluate the goodness-of-fit of neutrino oscillation data in the
framework of four-neutrino models. There this method was called {\it
  parameter goodness-of-fit} (PG), and it can be applied when the global
data consists of statistically independent subsets. The PG is based on
parameter estimation and hence it avoids the problem of being diluted
by many data points. It tests the {\it compatibility} of the different
data sets in the framework of the given theoretical model. In this
note we give a formal derivation of the probability distribution
function (\pdf) for the test statistic of the PG, and discuss the
application and interpretation of the PG on some examples. The
original motivation for the PG was the analysis of neutrino
oscillation data. However, the method may be very useful
also in other fields of physics, especially where global fits of many
parameters to data from several experiments are performed.

The outline of the paper is as follows. In Sec.~\ref{sec:gofs} we
define the PG and show that its construction is very similar to the
one of the standard goodness-of-fit. The formal derivation of the
\pdf\ for the PG test statistic is given in Sec.~\ref{sec:proof},
whereas in Sec.~\ref{sec:examples} a discussion of the application and
interpretation of the PG is presented. In Sec.~\ref{sec:correlations}
we consider the PG in the case of correlations due to theoretical errors,
and we conclude in Sec.~\ref{sec:conclusions}.

\section{Goodness-of-fit tests}
\label{sec:gofs}

We would like to start the discussion by citing the goodness-of-fit
definition given by the Particle Data Group (see Sec.~31.3.2.\ of
Ref.~\cite{pdg}):
``Often one wants to quantify the level of agreement between the data
and a hypothesis without explicit reference to alternative hypotheses.
This can be done by defining a {\it goodness-of-fit statistic}, $t$,
which is a function of the data whose value reflects in some way the
level of agreement between the data and the hypothesis. [\ldots] The
hypothesis in question, say, $H_0$ will determine the \pdf\ $g(t|H_0)$
for the statistic. The goodness-of-fit is quantified by giving the
$p$-value, defined as the probability to find $t$ in the region of
equal or lesser compatibility with $H_0$ than the level of
compatibility observed with the actual data. For example, if $t$ is
defined such that large values correspond to poor agreement with the
hypothesis, then the $p$-value would be
\begin{equation} \label{eq:pdfgof}
    p = \int_{t_\mathrm{obs}}^\infty g(t|H_0) {\rm d}t \,,
\end{equation}
where $t_\mathrm{obs}$ is the value of the statistic obtained in the
actual experiment.''

Let us stress that from this definition of goodness-of-fit one has
complete freedom in choosing a test statistic $t$, as long as the
correct \pdf\ for it is used.

\subsection{The standard goodness-of-fit}
\label{sec:defsgof}

Consider $N$ random observables $\bnu=(\nu_i)$ and let
$\mu_i(\btheta)$ denote the expectation value for the observable
$\nu_i$, where $\btheta = (\theta_\alpha)$ are $P$ independent
parameters which we wish to estimate from the data. Assuming that the
covariance matrix $S$ is known one can construct the following
$\chi^2$-function:
\begin{equation}\label{eq:def-chi2}
    \chi^2(\btheta) = [\bnu - \bmu(\btheta)]^T S^{-1} [\bnu - \bmu(\btheta)]
\end{equation}
and use its minimum $\chi^2_\mathrm{min}$ as test statistic for
goodness-of-fit evaluation:
\begin{equation}\label{eq:statsgof}
    t(\bnu) = \chi^2_\mathrm{min} \,.
\end{equation}
The hypothesis we want to test determines the \pdf\ $g(t)$ for this
statistics. Once the real experiments have been performed, giving the
results $\bnu_\obs$, the goodness-of-fit is given by the probability
of obtaining a $t$ larger than $t_\mathrm{obs}$, as expressed by
Eq.~\eqref{eq:pdfgof}. We will refer to this procedure as {\it
  standard goodness-of-fit} (\sgof):
\begin{equation} \label{eq:def-sgof}
    p_\sgof = \int_{\chi^2_\mathrm{min}(\bnu_\mathrm{obs})}^\infty
    g(t) \; {\rm d}t \,.
\end{equation}

The great success of this method is mostly due to a very powerful
theorem, which was proven over 100 years ago by
K.~Pearson\footnote{Pearson uses the slightly different test statistic
  \[
      \chi^2_{Pearson} = \sum_i
      \frac{[\nu_i - \mu_i(\btheta)]^2}{\mu_i(\btheta)}
  \]
  and assumes that the $\nu_i$ are independent. We prefer to use
  instead the $\chi^2$ of Eq.~\eqref{eq:def-chi2}, because in this way
  also correlated data can be considered.}~\cite{pearson} and which
greatly simplifies the task of calculating the integral in
Eq.~\eqref{eq:def-sgof}.
It can be shown under quite general conditions (see \eg\
Ref.~\cite{cramer}) that $\chi^2_\mathrm{min}$ follows a
$\chi^2$-distribution with $N-P$ degrees of freedom (\dof), so that
$g(t) = f_{\chi^2}(t,N-P)$. Therefore, the integral in
Eq.~\eqref{eq:def-sgof} becomes:
\begin{equation} \label{eq:simp-sgof}
    p_\sgof = \CL(\chi^2_\mathrm{min}(\bnu_\mathrm{obs}),\, N-P)
    \equiv \int_{\chi^2_\mathrm{min}(\bnu_\mathrm{obs})}^\infty
    f_{\chi^2}(t,N-P) \; {\rm d}t \,,
\end{equation}
where $\CL(\chi^2, n)$ is the {\it confidence level} function (see
\eg\ Fig.~31.1 of Ref.~\cite{pdg}).

In the following we propose a modification of the \sgof, for the case
when the data can be divided into several statistically independent
subsets.

\subsection{The parameter goodness-of-fit}
\label{sec:defpgof}

Consider $D$ statistically independent sets of random observables
$\bnu^r=(\nu^r_i)$ ($r=1,\ldots,D$), each consisting of $N_r$
observables ($i=1,\ldots,N_r$), with $N_\mathrm{tot} = \sum_r N_r$.
Now a theory depending on $P$ parameters $\btheta = (\theta_\alpha)$
is confronted with the data. The total $\chi^2$ is given by
\begin{equation}
    \chi^2_\mathrm{tot}(\btheta) = \sum_{r=1}^D \chi^2_r (\btheta) \,,
\end{equation}
where
\begin{equation}
    \chi^2_r(\btheta) =
    [\bnu^r - \bmu^r(\btheta)]^T S_r^{-1} [\bnu^r - \bmu^r(\btheta)]
\end{equation}
is the $\chi^2$ of the data set $r$. Now we define
\begin{equation}\label{eq:def-barchi2}
    \bar\chi^2(\btheta) =
    \chi^2_\mathrm{tot}(\btheta) - \sum_{r=1}^D \chi^2_{r,\mathrm{min}} \,,
\end{equation}
where $\chi^2_{r,\mathrm{min}} = \chi^2_r(\hat \btheta_r)$, and
$\hat\btheta_r(\bnu^r)$ are the values of the parameters which
minimize $\chi^2_r$. Instead of the total $\chi^2$-minimum we propose
now to use
\begin{equation}\label{eq:statpgof}
    t(\bnu) = \bar\chi^2_\mathrm{min} = \bar\chi^2(\tilde\btheta)
\end{equation}
as test statistic for goodness-of-fit evaluation. In
Eq.~\eqref{eq:statpgof} $\bar\chi^2_\mathrm{min}$ is the minimum of
$\bar\chi^2$ defined in Eq.~\eqref{eq:def-barchi2}, and $\tilde\btheta$
are the parameter values at the minimum of $\bar\chi^2$, or
equivalently of $\chi^2_\mathrm{tot}$. If we now denote by
$\bar{g}(t)$ the \pdf\ for this statistic, we can define the
corresponding goodness-of-fit by means of Eq.~\eqref{eq:pdfgof}, in
complete analogy to the SG case:
\begin{equation} \label{eq:def-pgof}
    p_\pgof = \int_{\bar\chi^2_\mathrm{min}(\bnu_\mathrm{obs})}^\infty
    \bar{g}(t) \; {\rm d}t \,.
\end{equation}
This procedure was proposed in Ref.~\cite{Maltoni:2002xd} with the
name \textit{parameter goodness-of-fit} (\pgof). Its construction 
is very similar to the \sgof, except that now $\bar\chi^2$ rather than
$\chi^2$ is used to define the test statistic.

In the next section we will show that also in the case of the PG the
calculation of the integral appearing in Eq.~\eqref{eq:def-pgof} can
be greatly simplified.  Let us define
\begin{equation}\label{eq:def-Pr}
P_r \equiv \mbox{rank}
\left[ \frac{\partial \bmu^r}{\partial \btheta} \right] \,.
\end{equation}
This corresponds to the number of {\it independent} parameters (or
parameter combinations), constrained by a measurement of
$\bmu^r$.\footnote{If in some pathological cases $P_r$ depends on the
point in the parameter space Eq.~\eqref{eq:def-Pr} should be evaluated
at the true values of the parameters, see Sec.~\ref{sec:proof_pgof}.}
Then under general condition $\bar\chi^2_\mathrm{min}$ is distributed
as a $\chi^2$ with $P_c = \sum_r P_r - P$ \dof, so that
Eq.~\eqref{eq:def-pgof} reduces to:
\begin{equation}\label{eq:simp-pgof}
    p_\pgof = \CL(\bar\chi^2_\mathrm{min}(\bnu_\mathrm{obs}),\, P_c).
\end{equation}

\section{The probability distribution function of $\bar\chi^2_\mathrm{min}$}
\label{sec:proof}

In this section we derive the distribution of the test statistic for
the \pgof. This can be done in complete analogy to the \sgof.
Therefore, we start by reviewing the corresponding proof for the
\sgof, see \eg\ Ref.~\cite{cramer}.

\subsection{The standard goodness-of-fit}
\label{sec:proof_sgof}

Let us start from the $\chi^2$ defined in Eq.~\eqref{eq:def-chi2}.
Since the covariance matrix $S$ is a real, positive and symmetric
matrix one can always find an orthogonal matrix $O$ and a diagonal
matrix $s$ such that $S^{-1} = O^T s^2 O$. Hence, we can write the
$\chi^2$ in the following way:
\begin{equation}\label{eq:chi2}
    \chi^2(\btheta) = [\bnu - \bmu(\btheta)]^T S^{-1} [\bnu - \bmu(\btheta)]
    = \by(\btheta)^T \by(\btheta) \,,
\end{equation}
where we have defined the new variables $\by(\btheta) = s O [\bnu -
\bmu(\theta)]$.  Let us denote the (unknown) true values of the
parameters by $\btheta^0$ and we define
\begin{equation}\label{eq:def-x}
    \bx \equiv \by(\btheta^0) = s O [\bnu - \bmu(\btheta^0)] \,.
\end{equation}
Now we assume that the $x_i$ are normal distributed with mean zero and
the covariance matrix ${\bf 1}_N$, which in particular implies that
they are statistically independent. This assumption is obviously
correct if the data $\nu_i$ are normal distributed with mean
$\mu_i(\btheta^0)$ and covariance matrix $S$.  However, it can be
shown (see \eg\ Refs.~\cite{cramer,frodesen,roe}) that this assumption
holds for a large class of arbitrary \pdf\ for the data under quite
general conditions, especially in the large sample limit, \ie\ large
$\nu_i$. Under this assumption it is evident that $\chi^2(\btheta^0) =
\bx^T\bx$ follows a $\chi^2$-distribution with $N$ \dof. According to
Eq.~\eqref{eq:statsgof} the test statistic $t$ for the \sgof\ is given
by the minimum of Eq.~\eqref{eq:chi2}. To derive the \pdf\ for $t$ we
state the following proposition:

\begin{propos} \label{prop:sgof}
    Let $\hat\btheta$ be the values of the parameters which minimize
    Eq.~\eqref{eq:chi2}. Then
    \begin{equation}
	\chi^2_\mathrm{min} =
	\chi^2(\theta^0) - \Delta\chi^2 \,,
    \end{equation}
    with $\chi^2_\mathrm{min} = \hat \by^T \hat \by$ and $\hat \by \equiv
    \by(\hat\btheta)$, has a $\chi^2$-distribution with $N-P$ \dof\ and
    $\Delta\chi^2$ has a $\chi^2$-distribution with $P$ \dof\ and is
    statistically independent of $\chi^2_\mathrm{min}$.
\end{propos}

A rigorous proof of this proposition is somewhat intricate and can be
found \eg\ in Ref.~\cite{cramer}. In the following we give an outline
of the proof dispensing with mathematical details for the sake of
clarity.

The $\hat\btheta$ are obtained by solving the equations
\begin{equation}\label{eq:mineqs}
    \frac{\partial \chi^2}{\partial \theta_\alpha}  =
    2 \by^T \frac{\partial \by}{\partial \theta_\alpha} = 0 \,.
\end{equation}
It can be proved (see \eg\ Ref.~\cite{cramer}) under very general
conditions that Eqs.~\eqref{eq:mineqs} have a unique solution
$\hat\btheta$ which converges to the true values $\btheta^0$ in the
large sample limit.  In this sense it is a good
approximation\footnote{Note that Eq.~\eqref{eq:approx} is exact if the
$\by$ depend linearly on the parameters $\btheta$.} to write
\begin{equation}\label{eq:approx}
    \hat\by \approx \bx + B (\hat\btheta - \btheta^0) \,,
\end{equation}
where we have defined the rectangular $N\times P$ matrix $B$ by
\begin{equation}\label{eq:def-B}
    B \equiv
    \LT. \frac{\partial \by}{\partial \btheta} \RT|_{\btheta^0} \,.
\end{equation}
With out loss of generality we assume that\footnote{If
  $\mathrm{rank}[B] = P' < P$ some of the parameters $\theta_\alpha$ are
  not independent. In this case one can perform a change of variables
  and choose a new sets of parameters $\theta'_\beta$, such that
  $\chi^2(\btheta')$ depends only on the first $P'$ of them. The
  remaining parameters are not relevant for the problem and can be
  eliminated from the very beginning. When repeating the construction in
  the new set of variables, the number of parameters will be equal to
  the rank of $B$.} $\mathrm{rank}[B] = P$. From Eq.~\eqref{eq:approx}
we obtain
\begin{equation}
    \LT. \frac{\partial \by}{\partial \btheta} \RT|_{\hat\btheta}
    \approx
    \LT. \frac{\partial \by}{\partial \btheta} \RT|_{\btheta^0} = B \,.
\end{equation}
Using this last relation in Eq.~\eqref{eq:mineqs} we find that $\hat \by$
fulfils $\hat \by^T B = 0$.
Multiplying Eq.~\eqref{eq:approx} from the left side by $B^T$ this
leads to
\begin{equation}\label{eq:mincond}
    B^T \bx = -B^T B (\hat\btheta - \btheta^0) \,.
\end{equation}
Using Eqs.~\eqref{eq:approx} and  \eqref{eq:mincond} we obtain
\begin{equation}\label{eq:yTy}
    \hat \by^T \hat \by = \bx^T \bx -
    (\hat\btheta - \btheta^0)^T B^T B (\hat\btheta - \btheta^0) \,.
\end{equation}
The symmetric $P\times P$ matrix $B^T B$ can be written as $B^T B = R
b^2 R^T$ with the orthogonal matrix $R$ and the diagonal matrix $b$,
and Eq.~\eqref{eq:mincond} implies $b^{-1} R^T B^T \bx = -b R^T
(\hat\btheta - \btheta^0)$. Defining the $N \times P$ matrix
\begin{equation}\label{eq:def-H}
    H \equiv B R b^{-1}
\end{equation}
we find $(\hat\btheta - \btheta^0)^T B^T B (\hat\btheta - \btheta^0) =
\bx^T H H^T \bx$, and Eq.~\eqref{eq:yTy} becomes
\begin{equation}\label{eq:yhat3}
    \hat \by^T \hat \by = \bx^T ({\bf 1}_N - HH^T) \bx \,.
\end{equation}
Note that the matrix $H$ obeys the orthogonality relation $H^T H =
{\bf 1}_P$, showing that the $P$ column vectors of length $N$ in $H$
are orthogonal. We can add $N-P$ columns to the matrix $H$ completing
it to an orthogonal $N\times N$ matrix: $V = (H, K)$. Here $K$ is an
$N \times (N-P)$ matrix with $K^T K = {\bf 1}_{(N-P)}$, $H^T K = 0$
and the completeness relation
\begin{equation}\label{eq:complete}
    V V^T = H H^T + K K^T = {\bf 1}_N \,.
\end{equation}
Now we transform to the new variables
\begin{equation}\label{eq:def-vw}
    \bx' = V^T \bx
    \;,\quad
    \bx' = \LT( \begin{array}{c} {\bf v} \\ {\bf w} \end{array} \RT) =
    \LT( \begin{array}{c} H^T \bx \\ K^T \bx \end{array} \RT) \,,
\end{equation}
where ${\bf v} = H^T \bx$ is a vector of length $P$ and ${\bf w} = K^T \bx$ is
a vector of length $N-P$. In general, if the covariance matrix of the random
variables $\bx$ is $S$, then the covariance matrix $S'$ of $\bx' = V^T \bx$ is
given by $S' = V^T S V$. Hence, since in the present case the $x_i$ are normal
distributed with mean 0 and covariance matrix ${\bf 1}_N$ the same is true for
the $x'_i$. In particular also ${\bf v}$ and ${\bf w}$ are statistically
independent. Using Eqs.~\eqref{eq:yhat3} and \eqref{eq:complete} we deduce
\begin{equation} \label{eq:chi2min}
    \hat \by^T \hat \by = \bx^T ({\bf 1}_N - HH^T) \bx
    = \bx^T K K^T \bx = {\bf w}^T {\bf w}
\end{equation}
proving that $\chi^2_\mathrm{min} = \hat \by^T \hat \by$ has a
$\chi^2$-distribution with $N-P$ \dof. Finally, we obtain
\begin{equation} \label{eq:deltachi2}
    \Delta\chi^2 = \chi^2(\theta^0) - \chi^2_\mathrm{min} =
    \bx^T  \bx - \hat\by^T \hat\by = \bx^T HH^T \bx
    = \bv^T \bv \,,
\end{equation}
showing that $\Delta\chi^2$ has a $\chi^2$-distribution with $P$ \dof\
and is statistically independent of $\chi^2_\mathrm{min}$. \hfill
$\Box$

\subsection{The parameter goodness-of-fit}
\label{sec:proof_pgof}

Moving now to the \pgof\ we generalize in an obvious way the formalism
of the previous section by attaching and index $r$ for the data set
to each quantity. We have
\begin{equation}
    \chi^2_\mathrm{tot} (\btheta) = \sum_r \by_r^T(\btheta) \by_r(\btheta)
    \,,\quad
    \chi^2_\mathrm{tot} (\btheta^0) = \sum_r \bx_r^T \bx_r \,,
\end{equation}
and
\begin{equation} \label{eq:def-chi2bar}
    \bar\chi^2(\btheta)
    \equiv  \chi^2_\mathrm{tot}(\btheta) - \sum_r \chi^2_\mathrm{r,min}
    = \sum_r \LT[ \by_r^T(\btheta) \by_r(\btheta) -
    \hat\by_r^T \hat\by_r \RT]  \,.
\end{equation}

\begin{propos} \label{prop:pgof}
    Let $\tilde\btheta$ be the values of the parameters which minimize
    $\bar\chi^2(\btheta)$, or equivalently
    $\chi^2_\mathrm{tot}(\btheta)$.  Then $\bar\chi^2_\mathrm{min} =
    \bar\chi^2(\tilde\btheta)$ follows a $\chi^2$-distribution with
    $P_c$ \dof, with
    \begin{equation}\label{eq:def-Br}
	P_c \equiv \mathcal{P} - P
	\,,\quad
	\mathcal{P} \equiv \sum_{r=1}^D P_r
	\,,\quad
	P_r \equiv \mathrm{rank}[B_r]
	\quad\mbox{and}\quad
	B_r \equiv
	\LT. \frac{\partial \by_r}{\partial \btheta} \RT|_{\btheta^0} \,.
    \end{equation}
\end{propos}

The matrices $B_r$ are of order $N_r \times P$. Since a given data set
$r$ may depend only on some of the $P$ parameters, or on some
combination of them, in general one has to consider the possibility of
$P_r \le P$.\footnote{Note that the definition of $P_r$ in
Eq.~\eqref{eq:def-Br} is equivalent to the one given in
Eq.~\eqref{eq:def-Pr}.} This means that the symmetric $P\times P$
matrix $B_r^T B_r$ can be writen as $R_r b_r^T b_r R_r^T$, where $R_r$
is an orthogonal matrix and $b_r$ is a $P_r\times P$ ``diagonal''
matrix, such that the diagonal $P\times P$ matrix $b_r^T b_r$ will
have $P_r$ non-zero entries. Let us now define the $P\times P_r$
``diagonal'' matrix $b_r^{-1}$ in such a way that $(b_r^{-1})_{ii}
\equiv 1/(b_r)_{ii}$ for each of the $P_r$ non-vanishing entries of
$b_r$, and all other elements are zero. In analogy to
Eq.~\eqref{eq:def-H} we introduce now the matrices
\begin{equation}\label{eq:def-Hr}
    H_r \equiv B_r R_r b_r^{-1} \,,
\end{equation}
which are of order $N_r \times P_r$. To prove
Proposition~\ref{prop:pgof} we define the vectors of length
$\mathcal{P}$
\begin{equation}
    {\bf Y}(\btheta) \equiv
    \LT(\begin{array}{c} H_1^T \by_1(\btheta) \\
    \vdots \\
    H_D^T \by_D(\btheta) \end{array}\RT)
    \,,\quad
    {\bf X} \equiv
    \LT(\begin{array}{c} H_1^T \bx_1 \\
    \vdots \\
    H_D^T \bx_D \end{array}\RT) =
    \LT(\begin{array}{c} \bv_1 \\ \vdots \\ \bv_D \end{array}\RT)\,.
\end{equation}

In the first part of the proof we show that $\bar\chi^2_\mathrm{min} =
\tilde{\bf Y}^T \tilde{\bf Y}$ with $\tilde{\bf Y} \equiv {\bf
  Y}(\tilde\btheta)$. With arguments similar to the ones leading to
Eq.~\eqref{eq:yTy} we find
\begin{equation}
    \sum_r \tilde \by_r^T \tilde \by_r =
    \sum_r \bx_r^T \bx_r -
    (\tilde\btheta - \btheta^0)^T \sum_r B_r^T B_r
    (\tilde\btheta - \btheta^0) \,.
\end{equation}
Using further Eq.~\eqref{eq:yhat3} for each $r$ we obtain
\begin{eqnarray}
    \bar\chi^2_\mathrm{min}
    &=&
    \sum_r \tilde\by_r^T \tilde\by_r - \sum_r \hat\by_r^T \hat\by_r
    \nonumber\\
    &=&
    \sum_r \bx_r^T H_r H_r^T \bx_r -
    (\tilde\btheta - \btheta^0)^T \sum_r B_r^T B_r
    (\tilde\btheta - \btheta^0) \,. \label{eq:barchi2min1}
\end{eqnarray}
On the other hand we can use that the minimum values
$\tilde\btheta$ are converging to the true values $\btheta^0$ in the
large sample limit and write $\tilde{\bf Y} \approx {\bf X} +
\mathcal{B} (\tilde\btheta - \btheta^0)$, where we have defined the
$\mathcal{P}\times P$ matrix
\begin{equation}
    \mathcal{B} \equiv
    \LT. \frac{\partial {\bf Y}}{\partial \btheta} \RT|_{\btheta^0}
    =
    \LT(\begin{array}{c} H_1^T B_1 \\ \vdots \\ H_D^T B_D \end{array}\RT)\,.
\end{equation}
Without loss of generality we assume that $\mathrm{rank}[\mathcal{B}]
= P$.  Again, with arguments similar to the ones leading to
Eq.~\eqref{eq:yTy} we derive
\begin{equation} \label{eq:YTY}
    \tilde{\bf Y}^T \tilde{\bf Y} =
    {\bf X}^T {\bf X} -
    (\tilde\btheta - \btheta^0)^T
    \mathcal{B}^T \mathcal{B}
    (\tilde\btheta - \btheta^0) \,.
\end{equation}
Using Eq.~\eqref{eq:def-Hr} it is easy to show that
$\mathcal{B}^T \mathcal{B} = \sum_r B_r^T B_r$, and by comparing
Eqs.~\eqref{eq:YTY} and \eqref{eq:barchi2min1} we can readily verify
the relation $\bar\chi^2_\mathrm{min} = \tilde{\bf Y}^T \tilde{\bf Y}$.

To complete the proof we identify ${\bf Y} \leftrightarrow \by$ and
${\bf X} \leftrightarrow \bx$ and proceed in perfect analogy to the
proof of Proposition~\ref{prop:sgof} given in
Sec.~\ref{sec:proof_sgof}.
In particular, from the arguments presented there it follows that the
elements of $\bv_r$ are $P_r$ independent Gaussian variables with mean
zero and variance one. Since the $D$ data sets are assumed to be
statistically independent the vector ${\bf X}$ contains $\mathcal{P}$
independent Gaussian variables with mean zero and variance one. In
analogy to the matrices $H,K$ of Sec.~\ref{sec:proof_sgof} we obtain
now the $\mathcal{P}\times P$ matrix $\mathcal{H}$ and the
$\mathcal{P}\times P_c$ matrix $\mathcal{K}$, which fulfil
$\mathcal{H H}^T + \mathcal{K K}^T = {\bf 1}_\mathcal{P}$, and
Eq.~\eqref{eq:YTY} becomes
\begin{equation}\label{eq:YTY2}
    \tilde{\bf Y}^T \tilde{\bf Y} =
    {\bf X}^T ({\bf 1}_\mathcal{P} - \mathcal{H H}^T) {\bf X} \,.
\end{equation}
In analogy to the vector ${\bf w}$ from Eq.~\eqref{eq:def-vw} we define
now ${\bf W} \equiv \mathcal{K}^T {\bf X}$, containing $P_c =
\mathcal{P} - P$ independent Gaussian variables with mean zero and
variance one, and Eq.~\eqref{eq:YTY2} gives
\begin{equation}\label{eq:barchi2min}
    \tilde{\bf Y}^T \tilde{\bf Y} =
    {\bf X}^T \mathcal{K}\mathcal{K}^T {\bf X} = {\bf W}^T {\bf W} \,.
\end{equation}
From Eq.~\eqref{eq:barchi2min} it is evident that
$\bar\chi^2_\mathrm{min} = \tilde{\bf Y}^T \tilde{\bf Y}$ follows a
$\chi^2$-distribution with $P_c$ \dof. \hfill $\Box$

Let us conclude this section by noting that both Proposition
\ref{prop:sgof} and \ref{prop:pgof} are {\it exact} if the data are
multi-normally distributed and the theoretical predictions $\bmu$,
$\bmu^r$ depend linearly on the parameters $\btheta$. If these
requirements are not fulfilled the simplified expressions
\eqref{eq:simp-sgof} and \eqref{eq:simp-pgof} are valid only
approximately, and to calculate the SG and the PG one should in
principle use the general formulas \eqref{eq:def-sgof} and
\eqref{eq:def-pgof} instead. However, we want to stress that under
rather general conditions $\chi^2_\mathrm{min}$ and
$\bar\chi^2_\mathrm{min}$ will be distributed as a $\chi^2$ {\it in
the large sample limit} (\ie\ for large $\bnu$ and $\bnu^r$,
respectively), so that even in the general case 
Eqs.~\eqref{eq:simp-sgof} and~\eqref{eq:simp-pgof} can still be used.

\section{Examples and discussion}
\label{sec:examples}

In this section we illustrate the application of the PG on some
examples. In Sec.~\ref{sec:ex1} we show that in the simple case of two
measurements of a single parameter the PG is identical to the
intuitive method of considering the difference of the two
measurements, and in Sec.~\ref{sec:ex2} we show the consistency of the
PG and the SG in the case of independent data points. In
Sec.~\ref{sec:ex3} we discuss the application of the PG to neutrino
oscillation data in the framework of a sterile neutrino scheme. This
problem was the original motivation to introduce the PG in
Ref.~\cite{Maltoni:2002xd}. In Sec.~\ref{sec:remarks} we add some
general remarks on the PG.

\subsection{The determination of one parameter by two experiments}
\label{sec:ex1}

Let us consider two data sets observing the data points $\bnu^1 =
(\nu^1_i)$ ($i=1,\ldots,N_1$) and $\bnu^2 = (\nu^2_i)$
($i=1,\ldots,N_2$). Further, we assume that the expectation values for
both data sets can be calculated from a theory depending on one
parameter $\eta$: $\bmu^r(\eta)$ ($r=1,2$), and all $\nu^r_i$ are
independent and normal distributed around the expectation values with
variance $\sigma^r_i$. Then we have the following $\chi^2$-functions
for the two data sets $r=1,2$:
\begin{equation}\label{eq:chi2ex}
    \chi^2_r(\eta) = \sum_{i=1}^{N_r} \LT(
    \frac{ \nu_i^r - \mu^r_i(\eta) } { \sigma^r_i }
    \RT)^2 =
    \chi^2_{r,\mathrm{min}} +
    \LT(
    \frac{ \hat\eta_r - \eta } { \hat\sigma_r }
    \RT)^2 \,,
\end{equation}
where $\hat\eta_r = \hat\eta_r(\bnu^r)$ is the value of the parameter
at the $\chi^2$-minimum of data set $r$. Now one may ask the question
whether the results of the two experiments are consistent. More
precisely, we are interested in the probability to obtain $\hat\eta_1$
and $\hat\eta_2$ under the assumption that both result from the {\it
  same} true value $\eta^0$.

A standard method (see \eg\ Ref.~\cite{frodesen} Sec.~14.3) to answer
this question is to consider the variable
\begin{equation}\label{eq:z}
    z = \frac{\hat\eta_1 - \hat\eta_2}
    {\sqrt{ \hat\sigma_1^2 + \hat\sigma_2^2 }}  \,.
\end{equation}
If the theory is correct $z$ is normal distributed with mean zero and
variance one. Hence we can answer the question raised above by citing
the probability to obtain $|z| \ge |z_\mathrm{obs}|$:
\begin{equation}\label{eq:pex}
    p = 1 - \int_{-|z_\mathrm{obs}|}^{|z_\mathrm{obs}|} f_N(z; 0, 1) {\rm d}z \,,
\end{equation}
where $f_N$ denotes the normal distribution.

If the PG is applied to this problem, one obtains from
Eq.~\eqref{eq:chi2ex}
\begin{equation}
    \bar\chi^2(\eta) =
    \LT(
    \frac{ \hat\eta_1 - \eta } { \hat\sigma_1 }
    \RT)^2 +
    \LT(
    \frac{ \hat\eta_2 - \eta } { \hat\sigma_2 }
    \RT)^2 \,,
\end{equation}
and after some simple algebra one finds $\bar\chi^2_\mathrm{min} =
z^2$, where $z$ is given in Eq.~\eqref{eq:z}. Obviously, applying
Eq.~\eqref{eq:simp-pgof} to calculated the $p$-value according to the
\pgof\ with the relevant number of \dof\ $P_c = 2 - 1 = 1$ leads to
the same result as Eq.~\eqref{eq:pex}.

Hence, we arrive at the conclusion that in this simple case of
testing the compatibility of two measurements for the mean of a
Gaussian, the \pgof\ is identical to the intuitive method of testing
whether the difference of the two values is consistent with zero.

\subsection{Consistency of PG and SG for independent data points}
\label{sec:ex2}

As a further example of the consistency of the PG method we consider
the case of $N$ statistically independent data points $\nu_i$. Let us
denote by $\sigma_i$ the standard deviation of the observation $\nu_i$
($i=1,\ldots,N$), and the corresponding theoretical prediction by
$\mu_i(\btheta)$, where $\btheta$ is the vector of $P$ parameters. For
simplicity, we assume that each of the $\mu_i$ depends at least on one
parameter. Then the $\chi^2$ is given by
\begin{equation}
    \chi^2(\btheta)
    = \sum_{i=1}^N \chi_i^2(\btheta) \,,
    \qquad \text{where} \qquad
    \chi_i^2(\btheta) = \frac{[\nu_i - \mu_i(\btheta) ]^2}{\sigma_i^2} \,,
\end{equation}
and from the SG construction (see Sec.~\ref{sec:proof_sgof}) we know
that $\chi^2_\mathrm{min}$ follows a $\chi^2$-distribution with $N-P$
degrees of freedom.
On the other hand, if we consider each single data point as an
independent data set and we apply the PG construction, we easily see
that $\chi_{i,\mathrm{min}}^2 = 0$ for each $i$. This implies
$\bar\chi^2(\btheta) = \chi^2(\btheta)$, and in particular
$\bar\chi^2_\textrm{min} = \chi^2_\textrm{min}$. Therefore, for the
specific case considered here one expects that SG and PG are
identical.

To show that this is really the case let us first note that each
matrix $\partial \mu_i / \partial \btheta$ consists just of \emph{a
single line}, and therefore it obviously has rank one. Hence,
Eq.~\eqref{eq:def-Pr} gives $P_i = 1$ for each $i$. This reflects the
fact that from the measurement of a single observable we cannot derive
\emph{independent} bounds on $P$ parameters, but only \emph{a single
combination} of them is constrained. Therefore, the number of \dof\
relevant for the calculation of the PG is given by $P_c = \sum_{i=1}^N
P_i - P = N - P$, which is exactly the number of \dof\
relevant for the SG. Hence, we have shown that in the considered case
the two methods are equivalent and consistent.

\subsection{Application to neutrino oscillation data}
\label{sec:ex3}

In this section we use real data from neutrino oscillation experiments
to discuss the application of the PG and to compare it to the SG.  We
consider the so-called (2+2) neutrino mass scheme, where a fourth
(sterile) neutrino is introduced in addition to the three standard
model neutrinos. In general this model is characterized by 9
parameters: 3 neutrino mass-squared differences $\dms$, $\dma$,
$\Delta m^2_\mathrm{LSND}$ and 6 mixing parameters
$\theta_\mathrm{sol}$, $\theta_\mathrm{atm}$, $\theta_\mathrm{LSND}$,
$d_\mu$, $\eta_s$, $\eta_e$. The interested reader can find precise
definitions of the parameters, applied approximations, an extensive
discussion of physics aspects, and references in
Refs.~\cite{Maltoni:2002xd,Maltoni:2001bc,Maltoni:2002ni}.  Here we
are interested mainly in the statistical aspects of the analysis, and
therefore we consider a simplified scenario.

We do not include LSND, KARMEN and all the experiments sensitive to
$\Delta m^2_\mathrm{LSND}$ and the corresponding mixing angle
$\theta_\mathrm{LSND}$. Hence, we are left with three data sets from
solar, atmospheric and reactor neutrino experiments. The solar data
set includes the current global solar neutrino data from the SNO,
Super-Kamiokande, Gallium and Chlorine experiments, making a total of 81
data points, whereas the atmospheric data sample includes 65 data
points from the Super-Kamiokande and MACRO experiments (for details of
the solar and atmospheric analysis see Ref.~\cite{Maltoni:2002ni}). In
the reactor data set we include only the data from the KamLAND and the
CHOOZ experiments, leading to a total of $13 + 14 = 27$ data
points~\cite{Maltoni:2002aw, Apollonio:2002gd}. In general the reactor
experiments (especially CHOOZ) depend in addition to $\dms$ and
$\theta_\mathrm{sol}$ also on $\dma$ and a further mixing parameter
$\eta_e$. However, we adopt here the approximation $\eta_e = 1$, which
is very well justified in the (2+2) scheme \cite{Maltoni:2001bc}. This
implies that the dependence on $\dma$ disappears and we are left with
the parameters $\dms$ and $\theta_\mathrm{sol}$ for both reactor
experiments, KamLAND as well as CHOOZ.

\begin{table}
    \begin{tabular}{|@{\quad}l@{\quad}|%
	  @{\quad}l@{\quad}|@{\quad}c@{\quad}c@{\quad}c@{\quad}|}
	\hline
	data set & parameters & $N$ & $\chi^2_\mathrm{min}/\dof$ & SG \\
	\hline
	reactor & $\dms,\theta_\mathrm{sol}$ & 27 & 11.5/25 & 99\% \\
	solar & $\dms,\theta_\mathrm{sol},\eta_s$ & 81 & 65.8/78 & 84\%  \\
	atmospheric & $\dma,\theta_\mathrm{atm},\eta_s, d_\mu$
	& 65 & 38.4/61 & 99\% \\
	\hline
    \end{tabular}
    \caption{\label{tab:datasets}%
      Parameter dependence, total number of data points,
      $\chi^2_\mathrm{min}$ and the corresponding SG for the three
      data sets.}
\end{table}

\begin{figure} \centering
   \includegraphics[width=0.7\linewidth]{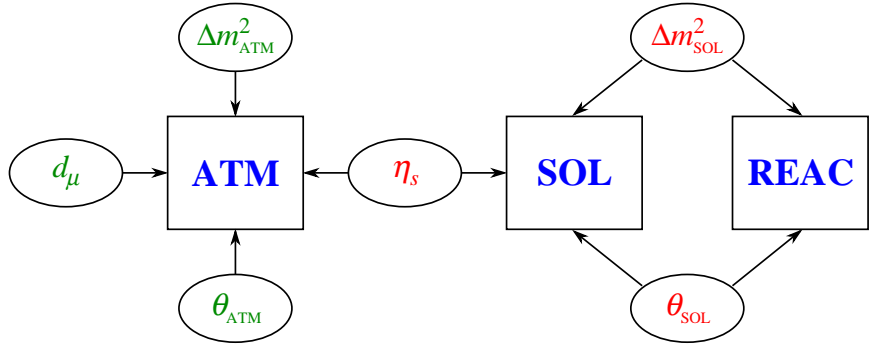}
   \caption{\label{fig:parameters} Parameter structure of the
      three data sets from reactor, solar and atmospheric neutrino
      experiments.}
\end{figure}

Under these approximations the experimental data sets we are using are
described only by the 6 parameters $\dms$, $\dma$,
$\theta_\mathrm{sol}$, $\theta_\mathrm{atm}$, $\eta_s$, $d_\mu$. The
parameter structure is illustrated in Fig.~\ref{fig:parameters}. This
simplified analysis serves well for discussing the statistical aspects
of the problem; a more general treatment including a detailed
discussion of the physics is given in
Refs.~\cite{Maltoni:2001bc,Maltoni:2002xd}. In Tab.~\ref{tab:datasets}
we summarize the parameter dependence, the number of data points, the
minimum values of the $\chi^2$-functions and the resulting SG. We
observe that all the data sets analyzed alone give a very good fit.
Let us remark that especially in the case of reactor and atmospheric
data the SG is suspicious high. This may indicate that the errors have
been estimated very conservatively.

\begin{figure} \centering
    \includegraphics[width=0.6\linewidth]{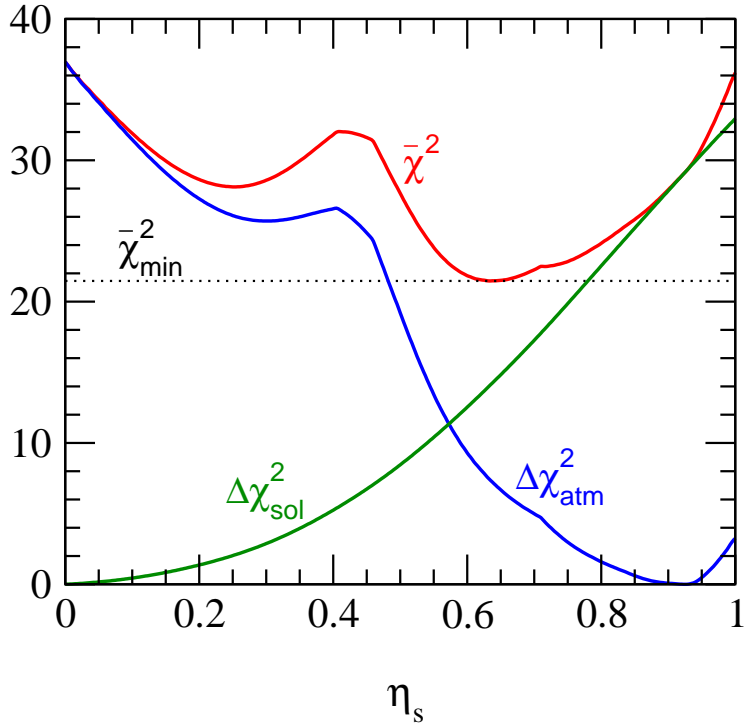}
    \caption{\label{fig:chi2} PG for solar and atmospheric neutrino data
      in the (2+2) scheme.}
\end{figure}

In Tab.~\ref{tab:PG} we show the results of an SG and PG analysis for
various combinations of the three data sets. In the first three lines
in the table only two out of the three data sets are combined. By
combining solar and atmospheric neutrino data we find a
$\chi^2_\mathrm{min}$ of 126.7. With the quite large number of \dof\
of 140 this gives an excellent SG of 78.3\%. If however, the PG is
applied we obtain a goodness-of-fit of only $3.54\times 10^{-6}$. The
reason for this very bad fit can be understood from
Figs.~\ref{fig:parameters} and \ref{fig:chi2}. From 
Fig.~\ref{fig:parameters} one
finds that solar and atmospheric data are coupled by the parameter
$\eta_s$. In Fig.~\ref{fig:chi2} the $\Delta \chi^2$ is shown for both
sets as a function of this parameter. We find that there is indeed
significant disagreement between the two data sets:\footnote{The
  physical reason for this is that both data sets strongly disfavour
  oscillations into sterile neutrinos. Since it is a generic prediction
  of the (2+2) scheme that the sterile neutrino must show up either in
  solar or in atmospheric neutrino oscillations the model is ruled out
  by the PG test~\cite{Maltoni:2002xd}.} solar data prefers values of
$\eta_s$ close to zero, whereas atmospheric data prefers values close
to one. There are two reasons why this strong disagreement does not
show up in the SG. First, since the SG of both data sets alone is very
good, there is much room to ``hide'' some problems in the combined
analysis. Second, because of the large number of data points many of
them actually might not be sensitive to the parameter $\eta_s$, where
the disagreement becomes manifest. Hence, the problem in the combined
fit becomes diluted due to the large number of data points. We
conclude that the PG is very sensitive to disagreement of the data
sets, even in cases where the individual $\chi^2$-minima are very low,
and when the number of data points is large.

\begin{table}
    \begin{tabular}{|@{\quad}l@{\quad}|%
	  @{\quad}c@{\quad}c@{\quad}c@{\quad}|%
	  @{\quad}c@{\quad}c@{\quad}c@{\quad}c@{\quad}|}
	\hline
	data sets & $N_\mathrm{tot}$ & $\chi^2_\mathrm{tot,min}/\dof$ & SG
	& $\sum_r P_r$ & $P$ & $\bar\chi^2_\mathrm{min}/P_c$ & PG \\
	\hline
	sol + atm   & 146 & 126.7/140 & 78.3\%
	& 3+4 & 6   & 21.5/1    & $3.54\times 10^{-6}$ \\
	react + sol & 108 & 77.4/105  & 98.0\%
	& 2+3 & 3   & 0.13/2    & 93.5\% \\
	react + atm & 92  & 49.9/86   & 99.9\%
	& 2+4 & 6   & 0.0/0     & $-$ \\
	\hline
	KamL + sol + atm  & 159 & 132.7/153 & 88.1\%
	& 2+3+4 & 6   & 21.7/3    & $7.53\times 10^{-5}$ \\
	react + sol + atm & 173 & 138.2/167 & 95.0\%
	& 2+3+4 & 6   & 21.7/3    & $7.53\times 10^{-5}$ \\
	\hline
    \end{tabular}
    \caption{\label{tab:PG}Comparison of SG and PG for various
      combinations of the data sets from solar, atmospheric and reactor
      neutrino experiments.}
\end{table}

In the reactor + solar analysis one finds complete agreement between
the two data sets for the SG as well as for the PG. This reflects the
fact that the determination of the parameters $\theta_\mathrm{sol}$
and $\dms$ from reactor and solar neutrino experiment are in excellent
agreement \cite{Maltoni:2002aw}. Finally, in the case of the combined
analysis of reactor and atmospheric data the PG cannot be applied.  In
our approximation these data sets have no parameter in common as one
can see in Fig.~\ref{fig:parameters}. Hence, it makes no sense to test
their compatibility, or even to combine them at all.

In the lower part of Tab.~\ref{tab:PG} we show the results from
combining all three data sets. By comparing these results with the one
from the solar + atmospheric analysis one can appreciate the advantage
of the PG. If we add only the 13 data points from KamLAND to the solar
and atmospheric samples we observe that the SG improves from 78.3\% to
88.1\%, whereas if both reactor experiments are included we
obtain an SG of 95.0\%. This demonstrates that the SG strongly depends on
the number of data points. Especially the 14 data points from CHOOZ
contain nearly no relevant information, since the best fit values of
$\dms$ and $\theta_\mathrm{sol}$ are in the no-oscillation regime for
CHOOZ implying that the $\chi^2$ is flat. Moreover, since
reactor data are not sensitive to the parameter $\eta_s$ (see
Fig.~\ref{fig:parameters}) the disagreement between solar and
atmospheric data becomes even more diluted by the additional
reactor data points. This clearly illustrates that the SG can be
drastically improved by adding data which contains no information on
the relevant parameters.
Also the PG improves slightly by adding reactor data, reflecting the
good agreement between solar and reactor data. However, the resulting
PG still is very small due to the disagreement between solar and
atmospheric data in the model under consideration. Moreover, the PG
is completely unaffected by the addition of the CHOOZ data, because the
$\chi^2$ of CHOOZ is flat in the relevant parameter region, and the PG
is sensitive only to the parameter dependence of the data sets.

Finally, we mention that in view of the analyses shown in
Tab.~\ref{tab:PG} the meaning of $P_c$, the number of \dof\ for PG
becomes clear. It corresponds to the number of parameters coupling the
data sets. Solar and atmospheric data are coupled only by $\eta_s$,
hence $P_c = 1$, whereas reactor and solar data are coupled by
$\theta_\mathrm{sol}$ and $\dms$ and $P_c = 2$. Atmospheric data has
no parameter in common with reactor data, therefore $P_c = 0$.  In the
combination of reactor + solar + atmospheric data sets the three
parameters $\eta_s, \theta_\mathrm{sol}, \dms$ provide the coupling and
$P_c = 3$.

\subsection{General remarks on the PG}
\label{sec:remarks}

({\it a}) Using the relation $\bar\chi^2_\mathrm{min} = \sum_r
\Delta\chi^2_r(\tilde\btheta)$ one can obtain more insight into the
quality of the fit by considering the contribution of each data set to
$\bar\chi^2_\mathrm{min}$.  If the PG is poor it is possible to
identify the data sets leading to the problems in the fit by looking
at the individual values of $\Delta\chi^2_r(\tilde\btheta)$. In this
sense the PG is similar to the so-called ``pull approach'' discussed
in Ref.~\cite{Fogli:2002pt} in relation with solar neutrino analysis.

({\it b}) One should keep in mind that the PG is completely
insensitive to the goodness-of-fit of the {\it individual} data sets.
Because of the subtraction of the $\chi^2_{r,\mathrm{min}}$ in
Eq.~\eqref{eq:def-barchi2} all the information on the quality of the
fit of the data sets alone is lost. One may benefit from this property
if the SG of the individual data sets is very good (see the example in
Sec.~\ref{sec:ex3}). On the other hand, if \eg\ one data set gives a
bad fit on its own this will not show up in the PG. Only the {\it
  compatibility} of the data sets is tested, irrespective of their
individual SGs.

({\it c}) The PG might be also useful if one is interested whether a
data set consisting of very few data points is in agreement with a
large data sample.\footnote{For example one could think of a
  combination of the 19 neutrino events from the Super Nova 1987A with
  the high statistics global solar neutrino data.} The SG of the
combined analysis will be completely dominated by the large sample and
the information contained in the small data sample may be drowned out
by the large number of data points. In such a case the PG can give
valuable information on the compatibility of the two sets, because it
is not diluted by the number of data points in each set and it is
sensitive only to the parameter dependence of the sets.

\section{Correlations due to theoretical uncertainties}
\label{sec:correlations}

One of the limitations of the PG is that it can be applied only if the
data sets are {\it statistically independent}. In many physically
interesting situations (for example, different solar neutrino
experiments) this is not the case since {\it theoretical
uncertainties} introduce correlations between the results of different
--~and otherwise independent~-- experiments. However, in such a case
one can take advantage of the so-called pull approach, which, as
demonstrated in Ref.~\cite{Fogli:2002pt}, is equivalent to the usual
covariance method. In that paper it was shown that if correlations due
to theoretical errors exist, it is possible to account for them by
introducing new parameters $\xi_a$ and adding penalty functions to the
$\chi^2$. In this way it is possible to get rid of unwanted
correlations and the PG can be applied.  The correlation parameters
$\xi_a$ should be treated in the same way as the parameters $\btheta$
of the theoretical model.

In this section we illustrate this procedure by considering a generic
experiment with an uncertainty on the normalisation of the predicted
number of events.  Let the experiment observe some energy spectrum
which is divided into $N$ bins. The theoretical prediction for the bin
$i$ is denoted by $\mu_i(\btheta)$ depending on $P$ parameters
$\btheta$. In praxis often $\mu_i(\btheta)$ is not known exactly. Let
us consider the case of a fully correlated relative error
$\sigma_\mathrm{th}$. A common method to treat such an error is to add
statistical and theoretical errors in quadrature, leading to the
correlation matrix
\begin{equation}\label{eq:correlation}
    S_{ij}(\btheta) = \delta_{ij} \sigma^2_{i,\mathrm{stat}} +
    \sigma^2_\mathrm{th} \mu_i(\btheta)\mu_j(\btheta) \,,
\end{equation}
where $\sigma_{i,\mathrm{stat}}$ is the statistical error in the bin
$i$. In the case of neutrino oscillation experiments such a correlated
error results \eg\ from an uncertainty of the initial flux
normalisation or of the fiducial detector volume. The $\chi^2$ is
given by
\begin{equation}\label{eq:chi2cov}
    \chi^2(\btheta) =
    \sum_{i,j=1}^N [\nu_i - \mu_i(\btheta)] S^{-1}_{ij}(\btheta)
    [\nu_j - \mu_j(\btheta)] \,,
\end{equation}
where $\nu_i$ are the observations.  As shown in
Ref.~\cite{Fogli:2002pt}, instead of Eq.~\eqref{eq:chi2cov} we can 
equivalently use
\begin{equation}\label{eq:chi2pull}
    \chi^2(\btheta,\xi)  =
    \sum_{i=1}^N  
    \LT(
    \frac{\nu_i - \xi \mu_i(\btheta)}{\sigma_{i,\mathrm{stat}}}
    \RT)^2
    +
    \LT(\frac{\xi -1}{\sigma_\mathrm{th}} \RT)^2 \,,
\end{equation}
and minimize with respect to the new parameter $\xi$.

On the other hand, if $\xi$ is considered as an additional parameter,
on the same footing as $\btheta$, all the data points
are formally uncorrelated and it is straight forward to apply the PG.
Subtracting the minimum of the first term in Eq.~\eqref{eq:chi2pull}
with respect to $\btheta$ and $\xi$ one obtains
\begin{equation}\label{eq:barchi2pull}
    \bar\chi^2(\btheta, \xi) =
    \Delta\chi^2(\btheta, \xi)
    +
    \LT(\frac{\xi -1}{\sigma_\mathrm{th}} \RT)^2 \,.
\end{equation}
The external information on the parameter $\xi$ represented by the
second term in Eq.~\eqref{eq:barchi2pull} is considered as an
additional data set. Evaluating the minimum of
Eq.~\eqref{eq:barchi2pull} for 1 \dof\ is a convenient method to test
if the best fit point of the model is in agreement with the constraint
on the over-all normalisation. In particular one can identify whether
a problem in the fit comes from the spectral shape (first term) or the
total rate (second term).

Moreover, one may like to divide the data into two parts, set I
consisting of bins $1,\ldots,n$ and and set II consisting of bins
$n,\ldots,N$, and test whether these data sets are
compatible. Eq.~\eqref{eq:chi2pull} can be written as
\begin{equation}
    \chi^2(\btheta, \xi) = \sum_{i=1}^n  
    \LT(
    \frac{\nu_i - \xi \mu_i(\btheta)}{\sigma_{i,\mathrm{stat}}}
    \RT)^2
    +
    \sum_{i=n}^N  
    \LT(
    \frac{\nu_i - \xi \mu_i(\btheta)}{\sigma_{i,\mathrm{stat}}}
    \RT)^2
    +
    \LT(\frac{\xi -1}{\sigma_\mathrm{th}} \RT)^2 \,,
\end{equation}
and subtracting the minima of the two first terms gives the
$\bar\chi^2$ relevant for the PG:
\begin{equation}\label{eq:barchi2pull2}
    \bar\chi^2(\btheta, \xi) =
    \Delta\chi^2_\mathrm{I}(\btheta, \xi) +
    \Delta\chi^2_\mathrm{II}(\btheta, \xi)
    +
    \LT(\frac{\xi -1}{\sigma_\mathrm{th}} \RT)^2 \,.
\end{equation}
Assuming that the data sets I and II both depend on all $P$ parameters
$\btheta$ the minimum of this $\bar\chi^2$ has to be evaluated for
$P+2$ \dof\ to obtain the PG. This procedure tests whether the data
sets I and II are consistent with each other {\it and} the constraint
on the over-all normalisation. By considering the relative
contributions of the three terms in Eq.~\eqref{eq:barchi2pull2} it is
possible to identify potential problems in the fit. For example one
may test whether a bad fit is dominated only by a small subset of the
data, \eg\ a few bins at the low or high end of the spectrum.
Alternatively, the two data sets I and II can come from two different
experiments correlated by a common normalization error, \eg\ two
detectors observing events from the same beam.

It is straight forward to apply the method sketched in this section
also in more complicated situations. For example, if there are several
sources of theoretical errors leading to more complicated correlations
the pull approach can also be applied by introducing a parameter
$\xi_a$ for each theoretical error~\cite{Fogli:2002pt}. In a similar
way one can treat the case when the compatibility of several
experiments should be tested, which are correlated by common
theoretical uncertainties. (Consider \eg, the various solar neutrino
experiments, which are correlated due to the uncertainties on the
solar neutrino flux predictions.)

\section{Conclusions}
\label{sec:conclusions}

In this note we have discussed a goodness-of-fit method
which was proposed in Ref.~\cite{Maltoni:2002xd}. The so-called {\it
  parameter goodness-of-fit} (PG) can be applied when the global data
consists of several statistically independent subsets.  Its
construction and application are very similar to the standard
goodness-of-fit. We gave a formal derivation of the probability
distribution function of the proposed test statistic, based on
standard theorems of statistics, and illustrated the application of
the PG on some examples. We have shown that in the simple case of two
data sets determining the mean of a Gaussian, the PG is identical to
the intuitive method of testing whether the difference of the two
measurements is consistent with zero. Furthermore, we have compared the
standard goodness-of-fit and the PG by using real data from neutrino
oscillation experiments, which have been the original motivation for
the PG. In addition we have illustrated that the so-called pull approach
allows to apply the PG also in cases where the data sets are
correlated due to theoretical uncertainties.

The proposed method tests the compatibility of different data sets,
and it gives sensible results even in cases where the errors are
estimated very conservatively and/or the total number of data points
is very large. In particular, it avoids the problem that a possible
disagreement between data sets becomes diluted by data points which
are insensitive to the problem in the fit. The PG can also be very
useful when a set consisting of a rather small number of data points
is combined with a very large data sample.

To conclude, we believe that physicists should keep an open mind when
choosing a statistical method for analyzing experimental data. In many
cases much more information can be extracted from data if the optimal
statistical tool is used. We think that the method discussed in this
note may be useful in several fields of physics, especially where
global analyses of large amount of data are performed.

\section*{Acknowledgment}

We thank C.~Giunti and P.~Huber for very useful discussions.
Furthermore, we would like to thank many participants of the NOON 2003
workshop in Kanazawa, Japan, for various comments on the PG. 
M.M.\ is supported by the Spanish grant BFM2002-00345, by the European
Commission RTN network HPRN-CT-2000-00148, by the ESF Neutrino
Astrophysics Network N.~86, and by the Marie Curie contract
HPMF-CT-2000-01008.
T.S.\ is supported by the ``Sonderforschungsbereich 375-95 f{\"u}r
Astro-Teilchenphysik'' der Deutschen Forschungsgemeinschaft.


\begin{thebibliography}{99}

\bibitem{pearson}
  K.~Pearson, Phil.\ Mag.\ V, 50 (1900) 157.

\bibitem{collins}
  J.C.~Collins and J.~Pumplin, arXiv:hep-ph/0105207.

\bibitem{Maltoni:2001bc}
  M.~Maltoni, T.~Schwetz and J.W.F.~Valle,
  Phys.\ Rev.\ D {\bf 65} (2002) 093004
  [arXiv:hep-ph/0112103].


\bibitem{solGof}
  P.~Creminelli, G.~Signorelli and A.~Strumia,
  JHEP {\bf 0105} (2001) 052 [arXiv:hep-ph/0102234];
  %
  M.V.~Garzelli and C.~Giunti,
  JHEP {\bf 0112} (2001) 017
  [arXiv:hep-ph/0108191].


\bibitem{Maltoni:2002xd}%
  M.~Maltoni, T.~Schwetz, M.A.~T{\`o}rtola, J.W.F.~Valle,
  Nucl.\ Phys.\ B {\bf 643} (2002) 321
  [arXiv:hep-ph/0207157].

\bibitem{pdg}
  K.~Hagiwara \textit{et al.}, [Particle Data Goup]
  Phys.\ Rev.\ D{\bf 66}, 010001-1 (2002).

\bibitem{cramer}
  H.~Cram\'er, \textit{Mathematical Methods of Statistics}, Princeton
  Univ. Press 1946.

\bibitem{frodesen}
  A.G.~Frodesen, O.~Skjeggestad and H.~Tofte, \textit{Probability and
    Statistics in Particle Physics}, Universitatsforlaget, Bergen 1979.

\bibitem{roe}
  B.P.~Roe, \textit{Probability and Statistics in Experimental Physics},
  Springer-Verlag, New York 1992.


\bibitem{Maltoni:2002ni}
  M.~Maltoni, T.~Schwetz, M.A.~T{\`o}rtola, J.W.F.~Valle,
  Phys.\ Rev.\ D {\bf 67} (2003) 013011
  [arXiv:hep-ph/0207227].

\bibitem{Maltoni:2002aw}
  M.~Maltoni, T.~Schwetz and J.W.F.~Valle,
  to appear in Phys.\ Rev. D, arXiv:hep-ph/0212129.

\bibitem{Apollonio:2002gd}
  M.~Apollonio {\it et al.},
  arXiv:hep-ex/0301017.

\bibitem{Fogli:2002pt}
  G.L.~Fogli, E.~Lisi, A.~Marrone, D.~Montanino and A.~Palazzo,
  Phys.\ Rev.\ D {\bf 66} (2002) 053010
  [arXiv:hep-ph/0206162].

\end{thebibliography}
\end{document}